\documentclass[showpacs,twocolumn]{revtex4}%
\usepackage{amsfonts}
\usepackage{amsmath}
\usepackage{amssymb}
\usepackage{graphicx}
\usepackage[dvips]{color}
\begin{document}

\title{Superconducting symmetry of three-dimensional  $t$-$J$ model on simple cubic lattice}
\author{Huai-Xiang Huang$^1$,YouQuan Li$^2$}

\affiliation {$^1$ Department of Physics, Shanghai University,
Shanghai, 200444,
China\\
$^2$ Zhejiang Institute of Modern Physics, Zhejiang University,
Hangzhou, 310027 , China }
\date{\today}

\date{\today}
\begin{abstract}
Motivated by the finding of nearly isotropic superconductivity in
$\mathrm{(Ba,K)Fe_2As_2}$, we use renormalized mean field theory
to investigated the $t$-$J$ model on three-dimensional simple
cubic lattice. A tunable anisotropic parameter is introduced to
dictate the coupling on $z$ direction. The symmetry of the
superconducting order is studied in detail. Calculation shows that
for the isotropic case, pairing parameters on the three
perpendicular directions have $\frac{2}{3}\pi$ phase shift to each
other. However, when the interaction on $z$ direction is
suppressed, the corresponding amplitude of the pairing parameter
decreases rapidly, furthermore, two-dimensional $d$-wave state
pairing is favored when the anisotropic rate less than $0.75$.
\end{abstract}

\maketitle
\section{Introduction}

The key feature of copper oxides is the layered structure and led
to speculation that reduced dimensionality is a necessary
prerequisite for superconductivity at temperatures above
$40K$~\cite{anderson}. Although two dimensional ($\mathrm{2D}$)
models, such as $t$-$J$ model or Hubbard model, have captured
essential of superconductivity, and successfully explained
properties of the un-doped insulator and occurrence of gap in
superconductors, $\mathrm{2D}$ models alone can not describe and
explain all observations of experiments~\cite{10,11,12,13,14,15}.
Despite how large the ratio between out-of-plane and in-plane
resistivity is, at the phase transition temperature $T_c$, both
resistivities drop to zero simultaneously, this indicates the
phase transition is of three dimensional~\cite{evidence},
meanwhile, the observed antiferromagnetism is definitely a
$3\mathrm{D}$ phenomenon~\cite{3d2}. Some works provided evidence
that superconductivity in the infinite-layer compounds
$\mathrm{ACuO_2}$~\cite{scaling} is of three-dimensional nature
which do not contain a charge-reservoir block and the distance
from one unit cell to the next is the shortest among all the
cuprates. Experimental data~\cite{10,11,12,13,14,15} shows that
the decrease of doping concentration is accompanied by a raise of
anisotropy which is defined as the ratio of the correlation
lengths parallel and perpendicular to the $\mathrm{CuO_2}$ plane
in cuprates superconductivity. For materials $\mathrm{Y_{123}}$,
$\mathrm{Y_{124}}$~\cite{y123,y124} and
$\mathrm{HgBa_2Ca_{0.86}Sr_{0.14}Cu_2O_{6-\delta}}$~\cite{scaling},
coherence length  and the anisotropy ratio imply that they are
anisotropic $3\mathrm{D}$ superconductors, these phenomena are
supported by good $3\mathrm{D}$ scaling analysis~\cite{scaling}.
Recently observed~\cite{chen,chen2,wang} superconductivity in
iron-arsenic-based compounds has attracted many attention. Our
motivation of investigating three dimensional isotropic
superconductivity come directly from the measurements of the
electrical resistivity in single crystals of
$\mathrm{(Ba,K)Fe_2As_2}$ in a magnetic field up to
$60\mathrm{T}$~\cite{hqyuan}, Yuan \textit{et al} found that the
superconducting properties are in fact quite isotropic, appear
more three dimensional than that of the copper oxides. Their
results indicates that reduced dimensionality in these compounds
is not necessarily a prerequisite for high temperature
superconductivity.

 $3\mathrm{D}$
anisotropic $t$-$J$ model has already been studied
before~\cite{compare,hole,T.schneider}. By using mean-field
Hamiltonian and carrying out expansion of free-energy, two main
results were obtained ~\cite{compare}: One is that transition
temperature decreases weakly with both increasing of $3\mathrm{D}$
coupling strength and doping concentration, the other one is that
in all cases $d$-wave pairing ansatz has the lowest energy.
However, in simple cubic lattice (SCL), each site has six nearest
neighbors(nn) settled in three perpendicular directions, no
direction is special, if superconducting behavior is possible, its
symmetry can not be conventional $d$-wave.

With the help of renormalized mean-field
theory\cite{zhang88,zhangtj}(RMFT), We found that for isotropic
case, pairing parameters on the three perpendicular directions
have $\frac{2}{3}\pi$ phase shift to each other. While as
interaction in $z$ direction is suppressed, corresponding
amplitude of pairing parameter drops quickly from infinite value
to zero. By tuning coupling integral in $z$ direction, our
calculation shows that superconducting symmetry are functions of
the anisotropic parameter and doping concentration. Moreover, as
the anisotropic parameter decreases from $1$ to $0.75$, symmetry
of pairing parameters change from $\frac{2}{3}\pi$ of
$3\mathrm{D}$ to $d$-wave of $2\mathrm{D}$. This may give some
understanding of $3\mathrm{D}$-$2\mathrm{D}$ crossover.

\section{ Formulation}
In SCL, anisotropic $t$-$J$ model can be written as $H=
P_dH_tP_d+H_s$ with
\begin{eqnarray}
H_t& =& -t\sum_{\langle nn\rangle\sigma}c^{\dag}_{i\sigma}c_{j\sigma} -t\lambda\sum_{\langle nn_{\perp}\rangle\sigma}c^{\dag}_{i\sigma}c_{j\sigma}+h.c. ,\nonumber\\
H_s& =& J\sum_{\langle nn\rangle}\vec{S}_{i}\cdot
\vec{S}_{j}+J\eta \sum_{\langle nn_{\perp}\rangle}\vec{S}_{i}\cdot
\vec{S}_{j},
\end{eqnarray}
where $P_d=\prod\limits_{i}(1- n_{i\uparrow }n_{i\downarrow })$ is
the Gutzwiller projection operator \cite{gutzwiller,H.Yokoyama}
which removes totally the doubly occupied states, $t$ and $J$ are
the electron hoping interaction and antiferromagnetic exchange
interaction, respectively. $c_{i\sigma }^{\dagger }$ is to create
an electron with spin $\sigma $ at site $i$, and $\vec{S}_{i}$ is
a spin operator. Summation $\langle nn\rangle$ runs over all nn in
$xy$ plane, while summation $\langle nn_{\perp}\rangle$ runs over
all nn in direction $z$ which is perpendicular to $xy$ plane. For
convenience, all anisotropic parameters are put into $z$ direction
terms, $\lambda$ and $\eta$ are anisotropic parameters with range
$[0,1]$, $\lambda=\eta=1$ corresponding to the isotropic case .

In RMFT the wavefunction of the Hamiltonian is assumed to be the
projected state $|\Psi\rangle=P_{d}|\Psi_{BCS}\rangle$,
$|\Psi_{BCS}\rangle=\prod_{\vec{k}}(u_{\vec{k}}+\upsilon_{\vec{k}}c^\dagger_{\vec{k}\uparrow}c^\dagger_{-{\vec{k}}\downarrow})|0\rangle$
, where $\vec{k}$ is constrained in the reduced Brillouin zone,
and the two coefficients satisfy $|u_k|^2+|\upsilon_k|^2=1$. The
projection operator can be taken into account by a set of
renormalized factors \cite{vollhardt,ogawa} defined as $\langle
c^{\dag}_{i\sigma}c_{j\sigma}\rangle\approx g_{t}\langle
c^{\dag}_{i\sigma}c_{j\sigma}\rangle_{0}$, $ \langle
\vec{S}_{i}\cdot \vec{S}_{j}\rangle \approx
g_{s}\langle\vec{S_i}\cdot \vec{S_j}\rangle_{0}$, where $\langle
\rangle_0$ denotes expectation value of unprojected state $\Psi_{
BCS}$, and $\langle \rangle$ denotes expectation value of physical
state $\Psi$. Then one has $\langle H \rangle= \langle H'
\rangle_0= \langle g_tH_t+g_sH_s \rangle_0$. In homogenous case
the renormalized factors \cite{zhang88,huang} take the form of
$g_{t}=2\delta /(1+\delta)$ and $g_{s}=4/(1+\delta)^2$.
Considering  even-parity case in which
\(u_{-\vec{k}}\upsilon^*_{-\vec{k}}=u_{\vec{k}}\upsilon^*_{\vec{k}}\)
and \(|\upsilon_{\vec{k}}|^2=|\upsilon_{-\vec{k}}|^2\), the
expectation value of the effective hamiltonian has the same form
as that of $2\mathrm{D}$
\begin{eqnarray}\label{h}
 \langle H' \rangle_0&=&2g_{t} \sum_{\vec{k}}\varepsilon_{\vec{k}}|\upsilon_{\vec{k}}|^2+N_{s}^{-1}\nonumber\\
 &
 &\times\sum_{\vec{k},\vec{k}'}V_{\vec{k},\vec{k}'}(|\upsilon_{\vec{k}}|^2|\upsilon_{\vec{k}'}|^2+u_{\vec{k}}\upsilon_{\vec{k}}^{*}\upsilon_{\vec{k}'}u_{\vec{k}'}^{*}),
\end{eqnarray}
where $N_s$ is the total number of sites and
\begin{eqnarray}\label{gm1}
\varepsilon_{\vec{k}}&=&-t\gamma^{\lambda}_{\vec{k}},\nonumber\\
V_{\vec{k}}&=&-\frac{3}{4} g_{s}J\gamma^{\eta}_{\vec{k}},\nonumber\\
\gamma^{\lambda}_{\vec{k}}&=&2(\cos{k_x}+\cos{k_y}+\lambda\cos{k_z}),\nonumber\\
\gamma^{\eta}_{\vec{k}}&=&2(\cos{k_x}+\cos{k_y}+\eta\cos{k_z}).
\end{eqnarray}

In order to investigate superconducting property, one should
introduce two mean-field parameters such as
particle-particle(pairing) parameter $\Delta_\tau=\langle
c^{\dag}_{i\uparrow}c^{\dag}_{i+\tau\downarrow}-c^{\dag}_{i\downarrow}c^{\dag}_{i+\tau\uparrow}\rangle_0
$ and particle-hole parameters $\xi_\tau=\sum_{\sigma}\langle
c^{\dag}_{i\sigma}c_{i+\tau,\sigma}\rangle_{0}$. By  minimizing
the quantity $W=\langle H'-\mu
\sum_{i\sigma}c_{i\sigma}^{\dagger}c_{i\sigma}\rangle_0$ with
respect to $u_{\vec{k}}$ and $\upsilon_{\vec{k}}$, where $\mu$ is
denoted as chemical potential, one gets the coupled gap equations
\begin{eqnarray}\label{gap equ2}
\Delta_\tau&=&N_{s}^{-1}\sum_{k}\cos{k_\tau} {\Delta}_{\vec{k}}/E_{\vec{k}},\\
\xi_\tau&=-&N_{s}^{-1}\sum_{k}\cos{k_\tau}\xi_{\vec{k}}/E_{\vec{k}},
\end{eqnarray}
where $\tau$ indicates the three perpendicular directions $x,y,z$,
$E_{\vec{k}}=\sqrt{\xi_{\vec{k}}^2+|\Delta_{\vec{k}}|^2}$,
$\Delta_{\vec{k}}= \Delta_{x}\cos{k_x}+\Delta_{y}\cos{k_y}+\eta
\Delta_z\cos{k_z}$,
$\xi_{\vec{k}}=\bar{\varepsilon}_k-\xi_{x}\cos{k_x}-\xi_{y}\cos{k_y}
-\eta\xi_z\cos{k_z}$,
$\bar{\varepsilon}_{\vec{k}}=(g_{t}\varepsilon_{\vec{k}}-\tilde{\mu})/(\frac{3}{4}g_{s}J)$,
and $\tilde{\mu}=\mu+N_{s} ^{-1}\langle \frac{\partial
H'}{\partial \delta}\rangle_0$. These gap equations should be
solved simultaneously with doping concentration
$\delta=N_{s}^{-1}\sum_{\vec{k}}\xi_{\vec{k}}/E_{\vec{k}}$. After
iterative self-consistent solving, for a set of given $\delta$,
$\lambda$, $\eta$ and $ \tilde{\mu}$ one can obtain all those
particle-particle and particle-hole parameters simultaneously.
Superconductivity symmetry is determined by the phase shift of
different pairing parameters $\Delta_{\tau}$ and the
superconductivity parameter $\Delta_{s\tau}$~\cite{zhang88} equals
$g_t\Delta_{\tau}$.
\begin{figure}
\includegraphics[width=8cm]{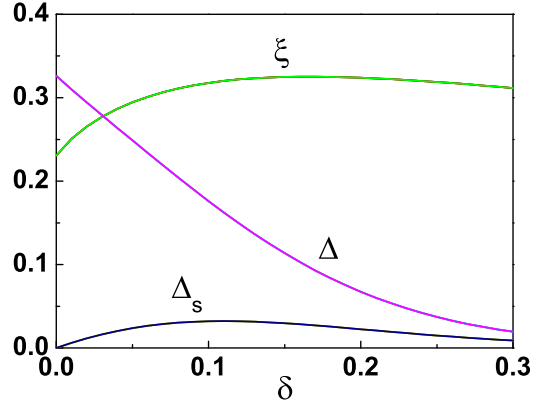}
\caption{\label{1}Parameters amplitude as functions of doping
concentration $\delta$. $\xi=\xi_i$ denotes amplitude of
particle-hole parameter, $\Delta=|\Delta_i|$ denotes pairing
parameter, and $ \Delta_s$ is superconducting parameter defined as
$g_t\Delta$.}
\end{figure}

\section{ symmetry of superconductivity for isotropic and anisotropic cases}
For isotropic SCL, $\eta=\lambda=1$. In half-filled case,
$\tilde\mu$, $\bar{\varepsilon}_k=0$, there is a trivial solution
with $\Delta_i=0$ corresponding to projected fermi-liquid.
$\xi_{\vec{k}}$ changes its sign at the surface and the average
energy of per site is $ \omega_f=-\frac{3J}{N_s}
\sum_{\xi_{\vec{k}},\xi_{\vec{k}'}<0}\gamma_{\vec{k}-\vec{k}'}\approx-0.505J
$. For non-trivial solution, by using Eq.(\ref{h}), the energy
per site can be written as $
 \omega=-\frac{3}{8}g_{s}J\sum_{\vec{k}}E_{\vec{k}}$, here relations
 $|u_{\vec{k}}|^2=\frac{1}{2}(1+\xi_{\vec{k}}/E_{\vec{k}})$ and
$u_{\vec{k}}\upsilon^*_{\vec{k}}= \frac{\Delta_{\vec{k}}}
{2E_{\vec{k}}}$ have been
 used.
By assuming $
E_{\vec{k}}=c(\cos{k_x}^2+\cos{k_y}^2+\cos{k_z}^2)^{1/2}$ and
substituting it into gap equations, one can get $
 c=\frac{1}{3N_s}\sum_{k}(\cos{k_x}^2+\cos{k_y}^2+\cos{k_z}^2)^{1/2}\approx0.398$ and the energy of per
site is $\omega=-\frac{9}{2}c^2J\approx-0.712J$, which is lower
than the energy of the projected fermi liquid state and is more
favored and stable. In the non-trivial case parameters should
satisfy following equations simultaneously
\begin{eqnarray}\label{phase}
&\,&\xi_{\tau}^2+| {\Delta}_\tau|^2=c^2,\nonumber\\
&\,&( {\Delta}_{\tau_1}
\Delta^{*}_{\tau_2}+h.c)+2\xi_{\tau_1}\xi_{\tau_2}=0.
\end{eqnarray}
It has $\mathrm{SU(2)}$ degeneracy, the most important solution is
\begin{eqnarray}\label{kinet largest}
\xi_\tau&=&\frac{\sqrt{3}}{3}c=0.229,\nonumber\\
|\Delta_\tau|&=&\frac{\sqrt{6}}{3}c=0.324,\nonumber\\
 {\Delta}_x&=&|\Delta|\exp{i\theta},\nonumber\\
 {\Delta}_y&=&|\Delta|\exp{i(\theta+2/3\pi)},\nonumber\\
 {\Delta}_z&=&|\Delta|\exp{i(\theta+4/3\pi)},
\end{eqnarray}
It clearly shows that the phase shift of different $\Delta_{\tau}$
is $\frac{2\pi}{3}$. By changing the sign of $\xi_y$ and taking
the phase difference of any two pairing parameters as
$|\frac{1}{3}\pi|$, one can obtain another solution and if one
sets one or two of the three $\xi$ as zero other solutions can
also be obtained. All these solutions have the same energy. Among
these energetically degenerated states the $\frac{2}{3}\pi$
symmetric state has the best kinetic energy $\langle H_t
\rangle_0$. Upon doping degeneracy will be lifted, superconducting
state favors the best kinetic energy state, which is the
$\frac{2\pi}{3}$ symmetry state. This is also the reason why we
call this solution as the most important one.

Hoping integral $t$ is used as energy unit, and  $t/J=3$ is taking
in order to be consistent with the superexchange relation of
$\mathrm{J}= 4\mathrm{t}^2/U$ in the large Hubbard $\mathrm{U}$
limit. For isotropic $\mathrm{SCL}$ case, self-consistent
parameters as functions of doping concentration are shown in
Fig.\ref{1}. Amplitude of all $\Delta_{\tau}$ are the same which
is denoted as $\Delta$ in the figure. Every $\xi_{\tau}$ is real
and has the same value of $\xi$. One can see from Fig.\ref {1}
that with doping increasing, amplitude of pairing parameters
decreases, while superconducting parameter $\Delta_s$ varies along
a non-monotonic curve. These properties are similar to that of
$2\mathrm{D}$ square lattice. The most interesting result is that
each $\Delta_{\tau}$ has imaginary part, $\theta_{\tau}$ is used
to denote phase of $\Delta_{\tau}$, the pairing parameters have
$\frac{2}{3}\pi$ phase shift to each other just as that of
half-filled case.

\begin{figure}
\includegraphics[width=8cm]{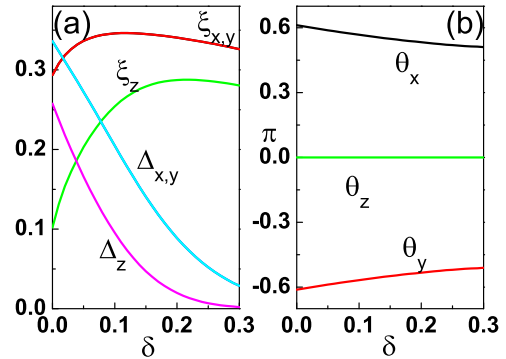}
\caption{Left picture shows mean-field parameters amplitude as
function of $\delta$ for $\eta=\lambda=0.9$. Right picture shows
how the phases $\theta_{\tau}$ varied as function of
$\delta$.}\label{2}
\end{figure}

\begin{figure}
\includegraphics[width=8cm]{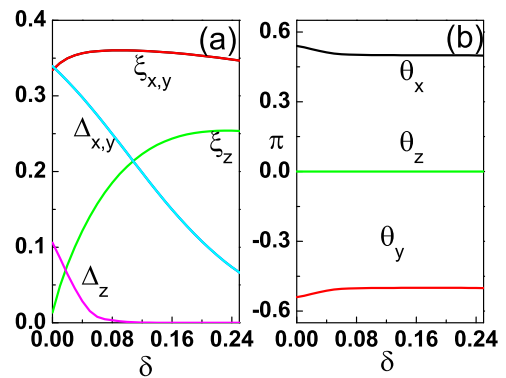}
\caption{Left picture shows mean-field parameters amplitude as
function of $\delta$ for $\eta=\lambda=0.8$. Right picture shows
how the phases $\theta_{\tau}$ varied as function of
$\delta$.}\label{3}
\end{figure}

When the interaction in $z$ direction is suppressed, amplitude of
the corresponding parameters will deviate from those of $xy$
plane. In order to make the situation more simpler, we set
$\eta=\lambda$. For $\eta=0.9$, the doping dependent parameters
are presented in Fig. \ref{2}. Fig. \ref{2}(a) shows that $\eta$
affects amplitudes of both paring parameter and particle-hole
parameter, with increasing $\delta$ all pairing parameters
decrease with $|\Delta_z|<|\Delta_{x,y}|$. Anisotropy also affects
the symmetry of the $\Delta_{\tau}$. Fig. \ref{2}(b) demonstrates
that at half-filled point
$\theta_{x,y}\equiv\theta_x-\theta_y<\frac{2}{3}\pi$, and
$\theta_{x,y}$ decreases with increasing $\delta$. Accompanied by
decrease of $|\Delta_z|$,  $\theta_{x,y}$ approaches to $\pi$. For
$\eta=0.8$, as shown in Fig. \ref{3}(a) with $\delta$ increasing
$|\Delta_z|$ drops more rapidly than $|\Delta_{x,y}|$ and vanishes
at $\delta=0.1$. Symmetry of pairing parameters are shown in
Fig.\ref{3}(b),  $\theta_{x,y}$ decreases from about $1.08\pi$ to
$\pi$ at $\delta=0.1$. For $\delta>0.1$, system apparently behaves
as $2\mathrm{D}$ with superconducting order being $d_{x^2-y^2}$
symmetry. By compare above two anisotropic cases one can
reasonably expect that at a given anisotropic parameter, system
will behave as $2\mathrm{D}$ in all doping level.

\begin{figure}
\includegraphics[width=8cm]{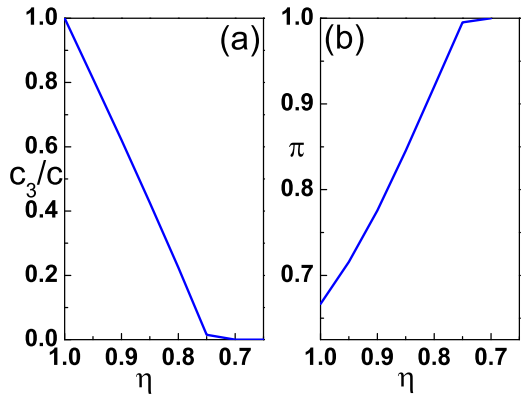}
\caption{For half-filled anisotropic case, with $\eta$ decreasing,
amplitude of $c_3=|\Delta_z|^2+\xi_z^2$ drops quickly. Phase
difference $\theta_{x,y}$ as function of anisotropic coefficient
is presented in right picture. }\label{4}
\end{figure}

This property can be demonstrated clearly in half-filled case. For
anisotropic half-filling,
$E_{\vec{k}}=\sqrt{c^2\cos^2{k_x}+c^2\cos^2{k_y}+c^2_3\eta^2\cos^2{k_z}}$
where $c^2=\xi^2_i+|\Delta|^2_i$, $c_3^2=\xi^2_z+|\Delta|^2_z$,
$i$ represents $x$ or $y$ direction. For a given $\eta$ one can
obtain the phase difference $\theta_{x,y}$ and the value of
$\frac{{c_3 }}{c}$ for the best kinetic energy state. As $c_3=c$,
it reduces to the isotropic case. As $c_3$ approaches to $0$,
degree of anisotropy is very large and the system turns to a
quasi-$2\mathrm{D}$ one. From Fig. \ref{4} one can see that by
decreasing $\eta$ from $1$, $\frac{c_3}{c}$ decreases quickly and
reaches zero at about $\eta=0.75$, simutanously the phase
difference $\theta_{x,y}$ increases from $\frac{2}{3}\pi$ to
$\pi$. These results indicate that as the anisotropic parameter
decreases to $0.75$, $\Delta_z$ vanishes, and system loses its
$3\mathrm{D}$ character.

\section{ Summary}
Experiment shows that $122$-type ternary iron arsenides possess
three-dimension properties~\cite{hqyuan}, although the microscopic
nature of superconductivity in iron-based compound is not clearly
at present, and one band model is not enough to describe them, we
investigated isotropic and anisotropic $t$-$J$ model on simple
cubic lattice to show the superconductivity symmetry  from
mean-field point of view. For isotropic three-dimensional $t$-$J$
model, superconductivity ground state is not conventional
$d$-wave, phase shift of each pairing parameter is exactly
$\frac{2}{3}\pi$. For anisotropic cases three-dimensional
character is not so obviously, adding a small anisotropic
interaction on $z$ direction will induces a great anisotropy in
its corresponding mean-field paring parameter and raise serious
instability of previous $3\mathrm{D}$ superconducting symmetry. We
found that pairing parameter $\Delta_z$ depends strongly on the
anisotropic parameter, as anisotropic parameter decrease to
$0.75$, system appears $2\mathrm{D}$ behavior. From this
discussion one can see that $3\mathrm{D}$ character superconductor
is sensitive to amplitude of couplings.

\section*{acknowledge}
The authors acknowledges professor Fu-Chun Zhang for  helpful
discussions during the research work. This work is supported by
NSF of China No.10747145, No.10874149 and by Shanghai Leading
Academic Discipline Project No. S30105.

\end{document}